\documentclass[aps,prd,twocolumn,showpacs,superscriptaddress]{revtex4-1}
\usepackage{amsmath}
\usepackage{latexsym}
\usepackage{amsfonts}
\usepackage{graphicx}
\usepackage{hyperref}

\begin{document}

\title{The challenge for single field inflation with BICEP2 result}

\author{Qing  Gao}
\email{gaoqing01good@163.com}
\affiliation{MOE Key Laboratory of Fundamental Quantities Measurement,
School of Physics, Huazhong University of Science and Technology,
Wuhan 430074,  P. R. China}

\author{Yungui Gong}
\email{yggong@mail.hust.edu.cn}
\affiliation{MOE Key Laboratory of Fundamental Quantities Measurement,
School of Physics, Huazhong University of Science and Technology,
Wuhan 430074,  P. R. China}
\affiliation{Institute of Theoretical Physics, Chinese Academy of Sciences, Beijing 100190, P. R. China}

\begin{abstract}
The detection of B-mode power spectrum by the BICEP2 collaboration constrains the tensor-to-scalar ratio
$r=0.20^{+0.07}_{-0.05}$ for the lensed-$\Lambda$CDM model. The consistency of this big value with the
{\em Planck} results requires a large running of the spectral index. The
large values of the tensor-to-scalar ratio and the running of the spectral index put a challenge
to single field inflation. For the chaotic inflation, the larger the value of the tensor-to-scalar ratio is,
the smaller the value of the running of the spectral index is. For the natural inflation, the absolute value of the
running of the spectral index has an upper limit.

\end{abstract}

\preprint{arXiv: 1403.5716}
\maketitle

\section{Introduction}

The detection of the primordial B-mode power spectrum by the BICEP2 collaboration confirms
the existence of primordial gravitational wave, and the observed B-mode power
spectrum gives the constraint on the tensor-to-scalar ratio
with $r=0.20^{+0.07}_{-0.05}$ at $1\sigma$ level for the lensed-$\Lambda$CDM model \cite{Ade:2014xna}.
Furthermore, $r=0$ is disfavored  at $7.0\sigma$ level. The new constraints on $r$
and the spectral index $n_s$ exclude a wide class of inflationary models.
For the inflation model with non-minimal coupling with gravity \cite{Kallosh:2013tua}, a universal
attractor at strong coupling was found with $n_s=1-2/N$ and $r=12/N^2$. This model is inconsistent
with the BICEP2 result $r\gtrsim 0.1$ at $2\sigma$ level
because the BICEP2 constraint on $r$ requires the number of e-folds $N=\sqrt{12/r}\lesssim \sqrt{120}\approx 11$
which is not enough to solve the horizon problem. If we require $N=50$,
then $r=0.0048$, so the model is excluded by the BICEP2 result. For the small-field inflation
like the hilltop inflation with the potential $V(\phi)=V_0[1-(\phi/\mu)^p]$ \cite{Albrecht:1982wi,Boubekeur:2005zm},
$r\sim 0$, so the model is excluded by the BICEP2 result.

Without the running of the spectral index, the combination of {\em Planck}+WP+highL data gives
$n_s=0.9600\pm 0.0072$ and $r_{0.002}<0.0457$ at the 68\% confidence level for
the $\Lambda$CDM model \cite{Ade:2013zuv,Ade:2013uln} which is in tension
with the BICEP2 result. When the running of the spectral index is included in the data fitting,
the same combination gives $n_s=0.957\pm 0.015$, $n_s'=dn_s/d\ln k=-0.022^{+0.020}_{-0.021}$
and $r_{0.002}<0.263$ at the 95\% confidence level \cite{Ade:2013zuv,Ade:2013uln}.
To give a consistent constraint on $r$
for the combination of {\em Planck}+WP+highL data and the BICEP2 data, we require
a running of the spectral index $n_s'<-0.002$ at the 95\% confidence level.
For the single field inflation, the spectral index $n_s$ for the
scalar perturbation deviates from the Harrison-Zel'dovich value of $1$ in the order of $10^{-2}$, so
$n_s'$ is in the order of $10^{-3}$. The explanation of large $r$ and $n_s'$ is a challenge to
single field inflation. In light of the BICEP2 data, several attempts were proposed to explain the large value
of $r$ \cite{Lizarraga:2014eaa,Harigaya:2014qza,Contaldi:2014zua,Collins:2014yua,Byrnes:2014xua,Anchordoqui:2014uua,Harigaya:2014sua,
Nakayama:2014koa,Zhao:2014rna,
Cook:2014dga,Kobayashi:2014jga,Miranda:2014wga,Masina:2014yga,Hamada:2014iga,Hertzberg:2014aha,
Dent:2014rga,Joergensen:2014rya,Freese:2014nla,Ashoorioon:2014nta,Ashoorioon:2013eia,
Choudhury:2014kma,Choudhury:2013iaa,Hotchkiss:2011gz,BenDayan:2009kv}.
In this Letter, we use the chaotic and natural inflation models to
explain the challenge.

\section{Slow-roll Inflation}

The slow-roll parameters are defined as
\begin{gather}
\label{slow1}
\epsilon=\frac{M_{pl}^2V_\phi^2}{2V^2},\\
\label{slow2}
\eta=\frac{M_{pl}^2V_{\phi\phi}}{V},\\
\label{slow3}
\xi=\frac{M_{pl}^4V_\phi V_{\phi\phi\phi}}{V^2},
\end{gather}
where $M^2_{pl}=(8\pi G)^{-1}$, $V_\phi=dV(\phi)/d\phi$, $V_{\phi\phi}=d^2V(\phi)/d\phi^2$
and $V_{\phi\phi\phi}=d^3V(\phi)/d\phi^3$. For the single field inflation, the spectral indices,
the tensor-to-scalar ratio and the running are
given by
\begin{gather}
\label{nsdef}
n_s-1\approx 2\eta-6\epsilon,\\
\label{rdef}
r\approx 16\epsilon\approx -8n_t,\\
\label{rundef}
n_s'=dn_s/d\ln k\approx 16\epsilon\eta-24\epsilon^2-2\xi.
\end{gather}
The number of e-folds before the end of inflation is given by
\begin{equation}
\label{efolddef}
N(t)=\int_t^{t_e}Hdt\approx \frac{1}{M_{pl}^2}\int_{\phi_e}^\phi\frac{V(\phi)}{V_\phi(\phi)}d\phi,
\end{equation}
where the value $\phi_e$ of the inflaton field at the end of inflation is defined by $\epsilon(\phi_e)=1$.
The scalar power spectrum is
\begin{equation}
\label{power}
\mathcal{P}_{\mathcal{R}}=A_s\left(\frac{k}{k_*}\right)^{n_s-1+n_s'\ln(k/k_*)/2},
\end{equation}
where the subscript ``*" means the value at the horizon crossing, the scalar amplitude
\begin{equation}
\label{power1}
A_s\approx \frac{1}{24\pi^2M^4_{pl}}\frac{\Lambda^4}{\epsilon}.
\end{equation}
With the BICEP2 result $r=0.2$, the energy scale of inflation is $\Lambda\sim 2.2\times 10^{16}$GeV.

\begin{figure}[htp]
\centerline{\includegraphics[width=0.45\textwidth]{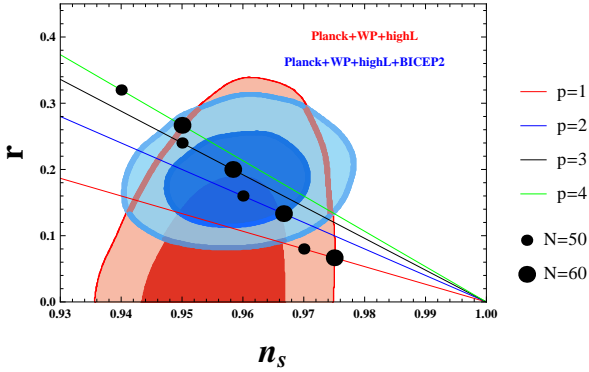}}
\caption{The $n_s-r$ diagrams for the chaotic inflation with $p=1$, $p=2$, $p=3$ and $p=4$. The 68\% and 95\% confidence contours
from the {\em Planck}+WP+highL data \cite{Ade:2013zuv,Ade:2013uln} and the {\em Planck}+WP+highL+BICEP2 data \cite{Ade:2014xna}
for the $\Lambda$CDM model are also shown.}
\label{pwrnsr}
\end{figure}

\begin{figure}[htp]
\centerline{\includegraphics[width=0.45\textwidth]{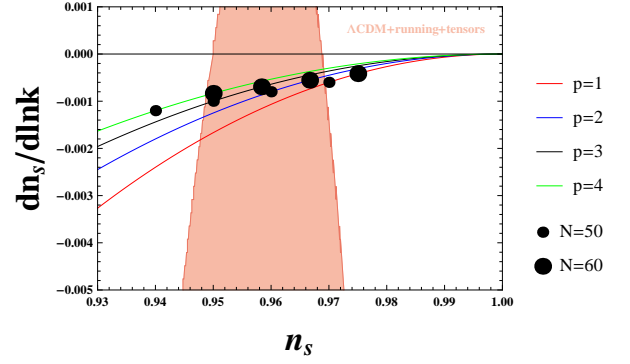}}
\caption{The $n_s-n_s'$ diagrams for the chaotic inflation with $p=1$, $p=2$, $p=3$ and $p=4$.
The 95\% confidence contour for the $\Lambda$CDM model
from the {\em Planck}+WP+highL data \cite{Ade:2013zuv,Ade:2013uln} is also shown.}
\label{pwrrun}
\end{figure}

For the chaotic inflation with the power-law potential $V(\phi)=\Lambda^4(\phi/M_{pl})^p$ \cite{linde83}, the slow-roll parameters
are $\epsilon=p/(4N_*)$, $\eta=(p-1)/(2N_*)$ and $\xi=(p-1)(p-2)/(4N^2_*)$. The spectral index $n_s=1-(p+2)/(2N_*)$,
the running of the spectral index $n_s'=-(2+p)/(2N_*^2)=-2(1-n_s)^2/(p+2)<0$ and the tensor-to-scalar ratio $r=4p/N_*=8p(1-n_s)/(p+2)$.
We plot the $n_s-r$ and $n_s-n_s'$ relations in Figs.
\ref{pwrnsr} and \ref{pwrrun} for $p=1$, $p=2$, $p=3$ and $p=4$. In Fig. \ref{pwrnsr}, we also show the points with
$N_*=50$ and $N_*=60$. From Figs. \ref{pwrnsr} and \ref{pwrrun}, we see that $r$ increases with the power $p$, but
$|n_s'|$ decreases with the power $p$. Therefore, it is not easy to satisfy both the requirements $r\gtrsim 0.1$ and
$n_s'<-0.002$. The chaotic inflation with $2<p<3$ is marginally consistent with the
observation at the 95\% confidence level.

\begin{figure}[htp]
\centerline{\includegraphics[width=0.45\textwidth]{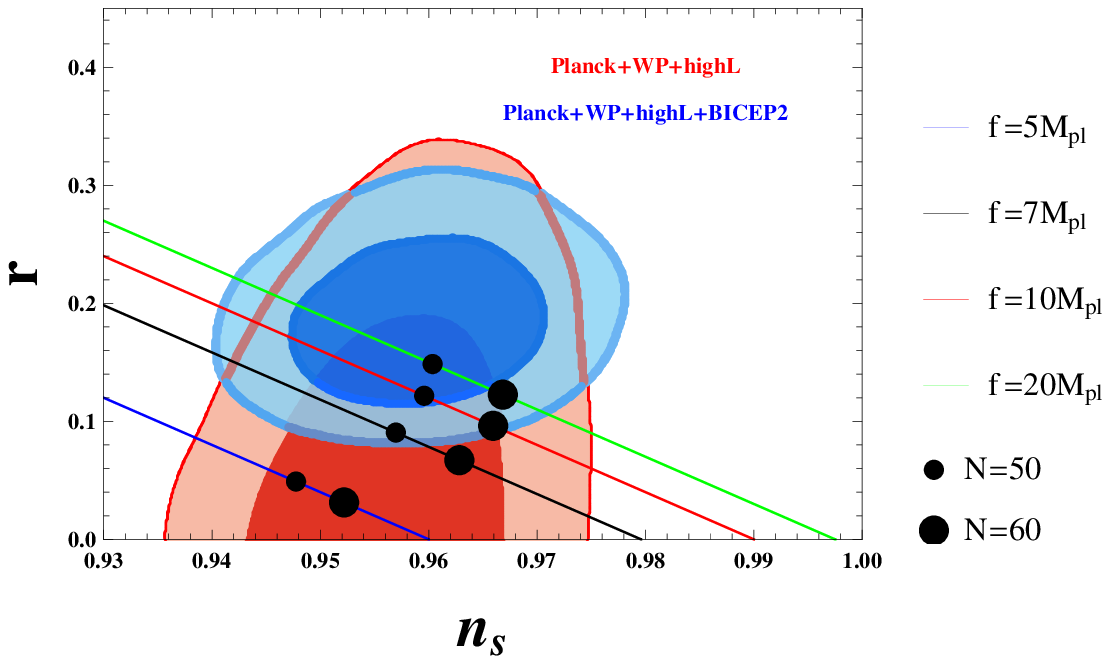}}
\caption{The $n_s-r$ diagrams for the natural inflation with $f=5M_{pl}$, $f=7M_{pl}$, $f=10M_{pl}$ and $f=20M_{pl}$.
The 68\% and 95\% confidence contours
from the {\em Planck}+WP+highL data \cite{Ade:2013zuv,Ade:2013uln} and the {\em Planck}+WP+highL+BICEP2 data \cite{Ade:2014xna}
for the $\Lambda$CDM model are also shown.}
\label{pngbnsr}
\end{figure}

\begin{figure}[htp]
\centerline{\includegraphics[width=0.45\textwidth]{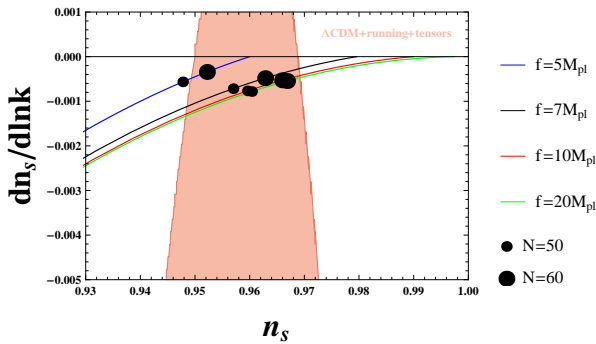}}
\caption{The $n_s-n_s'$ diagrams for the natural inflation with $f=5M_{pl}$, $f=7M_{pl}$, $f=10M_{pl}$ and $f=20M_{pl}$.
The 95\% confidence contour for the $\Lambda$CDM model
from the {\em Planck}+WP+highL data \cite{Ade:2013zuv,Ade:2013uln} is also shown.}
\label{pngbrun}
\end{figure}

For the natural inflation with the potential $V(\phi)=\Lambda^4[1+\cos(\phi/f)]$ \cite{Freese:1990rb}, the slow-roll parameters are
\begin{gather}
\label{pngbsl1}
\epsilon=\frac{M^2_{pl}}{2f^2}\left[\frac{\sin(\phi/f)}{1+\cos(\phi/f)}\right]^2,\\
\label{pngbsl2}
\eta=-\frac{M^2_{pl}}{f^2}\frac{\cos(\phi/f)}{1+\cos(\phi/f)},\\
\label{pngbsl3}
\xi=-\frac{M^4_{pl}}{f^4}\left[\frac{\sin(\phi/f)}{1+\cos(\phi/f)}\right]^2=-\frac{2M^2_{pl}}{f^2}\epsilon.
\end{gather}
Inflation ends when $\epsilon\sim 1$, so
\begin{equation}
\label{pngbphie}
\frac{\phi_e}{f}=\arccos\left[\frac{1-2(f/M_{pl})^2}{1+2(f/M_{pl})^2}\right],
\end{equation}
and the number of e-folds before the end of inflation is
\begin{equation}
\label{pngbn}
N_*=\frac{2f^2}{M^2_{pl}}\ln\left[\frac{\sin(\phi_e/2f)}{\sin(\phi_*/2f)}\right].
\end{equation}
Combining Eqs. (\ref{nsdef})-(\ref{rundef}) with (\ref{pngbsl1})-(\ref{pngbsl3}), we
plot the $n_s-r$ and $n_s-n_s'$ relations for the natural inflation with $f=5M_{pl}$, $f=7M_{pl}$,
$f=10M_{pl}$ and $f=20M_{pl}$ in Figs. \ref{pngbnsr} and \ref{pngbrun}. In Fig. \ref{pngbnsr},
we also show the points with $N_*=50$ and $N_*=60$. The results show that both $r$ and $|n_s'|$ increase
with the global symmetry breaking scale $f$. However, there is an upper limit on $|n_s'|$
which is only marginally consistent with the observation at the 95\% confidence level.
When $f/M_{mp}\gg 1$, the potential can be approximated by $V(\phi)=\Lambda^4(\phi/f-\pi)^2/2$
which is the power-law potential with $p=2$, this is the reason for the upper limit on $n_s'$.

\section{Conclusions}

For a single inflaton field with slow-roll, the tensor-to-scalar ratio $r\approx 16\epsilon$ which is linear
with the slow-roll parameter $\epsilon$, but the running of the spectral index $n_s'$ depends on the second
order slow-roll parameters, so $n_s'$ is at most in the order of $10^{-3}$. The BICEP2 and the {\em Planck}
data constrain $n_s'=-0.0221^{+0.011}_{-0.0099}$
and $r=0.20^{+0.07}_{-0.05}$ at the $1\sigma$ confidence level. Both the chaotic and natural inflation
are inconsistent with the observation at the $1\sigma$ level. The chaotic inflation with $2<p<3$
and the natural inflation with $f\gtrsim 10M_{pl}$ are
marginally consistent with the observation at the 95\% confidence level.
In conclusion, it is a challenge to simultaneously explain $r$ as large as $0.2$ and $n_s'$ as large as $-0.01$
for single field inflation. Unless the {\em Planck} and the BICEP2 data can be reconciled without
large $n_s'$, the challenge to single field inflation remains.

\begin{acknowledgments}
This work was partially supported by
the National Basic Science Program (Project 973) of China under
grant No. 2010CB833004, the NNSF of China under grant No. 11175270,
the Program for New Century Excellent Talents in University under grant No. NCET-12-0205
and the Fundamental Research Funds for the Central Universities under grant No. 2013YQ055.
\end{acknowledgments}


\end{document}